\documentclass[pra,aps,reprint,showpacs,amsmath,superscriptaddress]{revtex4-1}

\usepackage{bm}
\usepackage{graphicx}
\usepackage{dsfont}

\usepackage{mathrsfs}

\usepackage{framed}

\usepackage[all]{xy}
\usepackage{graphicx}
\usepackage{framed}
\usepackage{comment}
\usepackage{here}                    
\usepackage{latexsym}		     
\usepackage{amsmath}     
\usepackage{amsfonts}  
\usepackage{amssymb}
\usepackage{amsthm}
\usepackage{dcolumn}
\usepackage{color}
\usepackage{mathrsfs}

\usepackage{braket}
\usepackage{verbatim}

\newcommand{\be}{\begin{equation}}
\newcommand{\ee}{\end{equation}}
\newcommand{\bea}{\begin{eqnarray}}
\newcommand{\eea}{\end{eqnarray}}

\newcommand{\aver}[1]{\langle #1 \rangle}

\usepackage[colorlinks,citecolor=blue,linkcolor=blue,urlcolor=blue]{hyperref}

\begin{document}

\title{Intercept-Resend Emulation Attacks Against a Continuous-Variable Quantum Authentication Protocol with Physical Unclonable Keys}

\author{Lukas Fladung}
\affiliation{Institut f\"ur Angewandte Physik, Technische Universit\"at Darmstadt, D-64289 Darmstadt, Germany}

\author{Georgios M. Nikolopoulos}
\affiliation{Institut f\"ur Angewandte Physik, Technische Universit\"at Darmstadt, D-64289 Darmstadt, Germany}
\affiliation{Institute of Electronic Structure \& Laser, FORTH, P.O. Box 1385, GR-70013 Heraklion, Greece}

\author{Gernot  Alber}
\affiliation{Institut f\"ur Angewandte Physik, Technische Universit\"at Darmstadt, D-64289 Darmstadt, Germany}

\author{Marc Fischlin} 
\affiliation{Cryptoplexity, Technische Universit\"at Darmstadt, D-64289 Darmstadt, Germany}

\date{\today}

\begin{abstract}
Optical physical unclonable keys  are currently considered  to be rather promising candidates for the development of entity authentication protocols, which offer security against both classical and quantum adversaries. In this work we investigate the robustness of a continuous-variable  protocol, which relies on the scattering of coherent states of light from the key, against three different types of intercept-resend emulation attacks. The performance of the protocol is analysed for a broad range of physical parameters, and our results are compared to existing security bounds.
\end{abstract}


\maketitle


\section{Introduction}
\label{sec1}

Entity authentication (or identification) is a fundamental cryptographic task, which aims at providing a verifier with assurance about the identity of another entity (a claimant) \cite{handbook}.  In order to offer high levels of security, modern protocols combine in the framework of a challenge-response mechanism, something that the claimant possesses together with something that the claimant knows \cite{handbook2}. For instance, a typical transaction through an automatic-teller machine (verifier), 
relies on the smart card that the user possesses, and a secret personal identification number (PIN). 
In each transaction, the PIN is used to verify the user to the smart card, while the latter is equipped with a chip which runs a publicly-known  algorithm, and involves a numerical secret key on the card for which the verifier has a matching counterpart i.e., either shares the secret key with the verifier or holds the public key to the secret key.  Verification of the smart card is performed only if the PIN is correct, and involves a number of random and independent numerical challenges,  for which the chip on the card computes a response based on the implemented algorithm and the key. The user is authenticated only if the responses agree with the ones expected by the verifier. Attacks against conventional smart cards and  dynamic entity-authentication protocols  (EAP) are difficult but not impossible (e.g., see  \cite{PUFtutorial2014,holcomb2014pufs,Lim-etal:2005} and references therein). More precisely, physical invasive and non-invasive attacks, as well as software attacks (e.g., viruses), constitute a severe threat as they allow adversaries to extract the secret key from the card.  
The necessity for entity-authentication protocols that are robust against such types  of attacks, has motivated the development of physical unclonable functions  (PUFs) \cite{Pappu02,handbook3,PUFtaxonomy:2019}.

In general terms, a PUF is a cryptographic primitive which converts an input challenge into an output response, by means of a random physical unclonable key (PUK). The 
randomness of the PUK is introduced explicitly or implicitly during its fabrication, and it is considered to be technologically hard to clone (hence the term physical unclonable). The physical mechanism underpinning the operation of a PUF, as well as the nature of the challenge, depend on the nature of the PUK. Hence, one can have electronic, optical, magnetic, biological PUFs, etc. 
For a rather extensive list of PUFs and their classification the reader may refer to Ref.  \cite{PUFtaxonomy:2019}. So far, electronic and optical PUFs appear to be the most prominent classes of PUFs.  An advantage of the former is that they are compatible with existing technology and hardware, but their robustness against various types of modelling attacks does not seem to be as strong as expected (see \cite{cryptoeprint:2019:566,khalafalla2019pufs,ganji2019pufmeter,delvaux2019machine} and references therein). 
On the contrary, optical PUFs are not fully compatible with existing technology, but they offer many advantages relative to certain types of electronic PUFs, including high complexity, security against modeling attacks, and low cost \cite{opticalPUF2013}.

\begin{figure*}
\centerline{\includegraphics[width=0.8\linewidth]{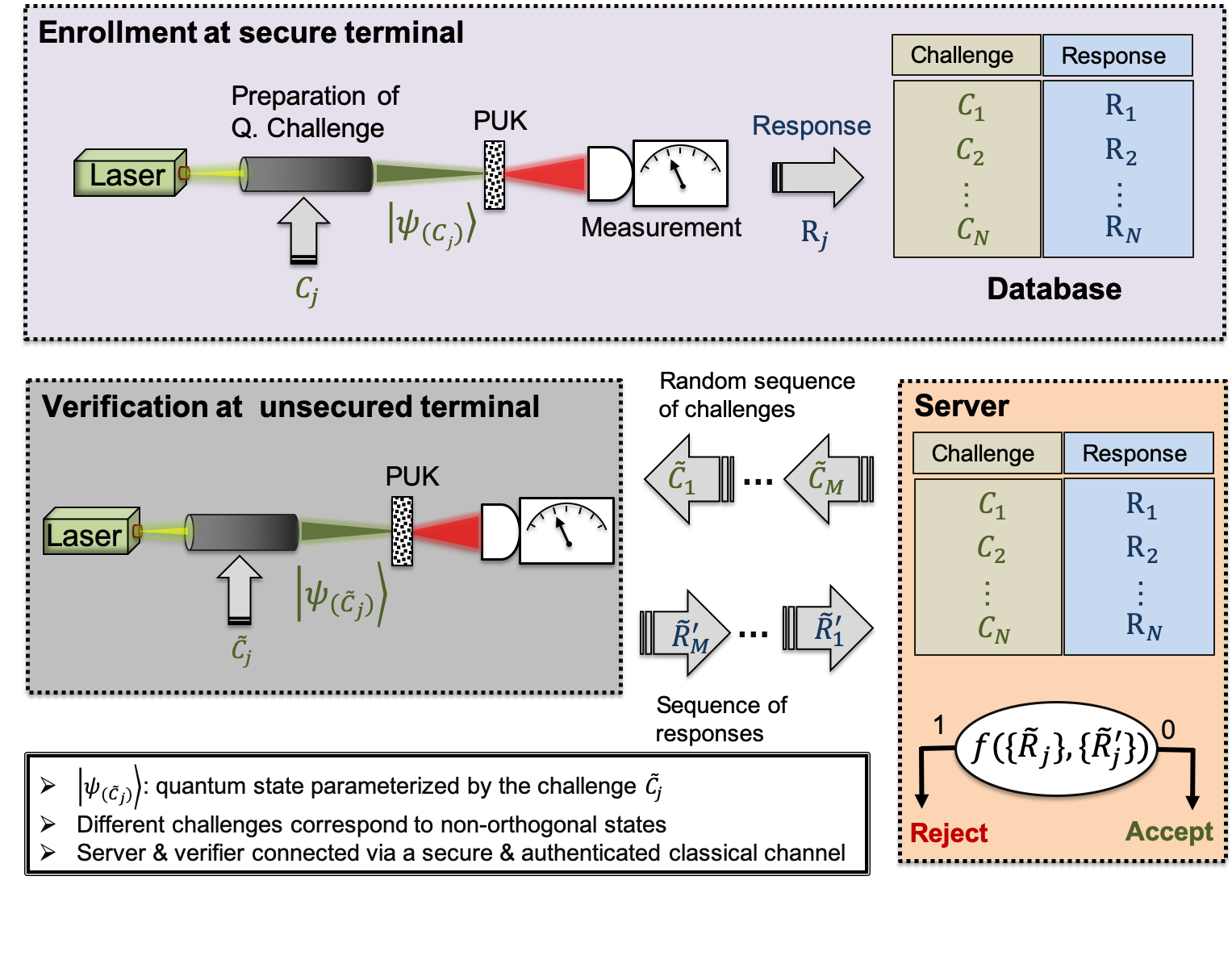}}
\caption{General schematic representation of the main stages and the typical operations of an EAP with quantum readout of a PUK. 
The enrollment stage is performed before the PUK is given to the user, and aims at the generation of a set of numerical CRPs. 
In the verification stage, $M\gg 1$ numerical challenges are chosen at random and independently from the available set of CRPs for the particular PUK. Each numerical challenge is encoded independently in the quantum state of a laser pulse, which is scattered from the PUK. The scattered light is measured and the outcome (response) is returned to the server. The PUK is accepted or rejected based on a publicly known function, which quantifies the deviations of the recorded responses from the expected ones.}
\label{qpuf_scheme:fig}
\end{figure*}

Another advantage of optical PUFs is that they accept quantum states of light as challenges, thereby enabling for the design of EAPs whose security relies on fundamental principles of quantum physics \cite{Goorden:14,nikolopoulos2017continuous}. Typically, the operation of such an EAP relies 
on a set of numerical challenge-response pairs (CRPs), which is generated during the enrollment stage, and it is stored in a secure database (see Fig. \ref{qpuf_scheme:fig}). There is also a publicly known bijective map of the set of numerical challenges onto a set of non-orthogonal quantum states of light. Whenever the holder of the PUK has to authenticate himself in a potentially unsecured  verification set-up, his PUK undergoes a verification procedure, where it is interrogated by a random sequence of 
non-orthogonal quantum states of light. Each state corresponds to a random and independently chosen numerical challenge. The optical response of the PUK to each quantum challenge is measured and processed so that to obtain a numerical response. The user is authenticated if the recorded numerical responses do not deviate considerably from the expected numerical responses listed in the set of CRPs. The  precise quantification of the deviations depends strongly on the details of the protocol, and plays a pivotal role in its security. The main point is that even when an adversary knows the set of numerical  CRPs to be used for the verification of a PUK,  
the impersonation of the legitimate user requires interaction of the adversary with the quantum states used for the interrogation of the PUK. This is the only way for the adversary to estimate the numerical challenges encoded on the quantum states, and to send the corresponding responses to the verifier by looking the estimated challenges up in the set of CRPs. However, fundamental theorems of quantum physics prevent perfect discrimination between non-orthogonal quantum states. 
 As a result, the intervention of the adversary will inevitably  introduce errors, that will be detected by the verifier, who can abort the authentication process.

In this context, we proposed recently a continuous-variable quantum authentication of optical PUKs, which relies on standard wavefront-shaping and homodyne-detection techniques \cite{nikolopoulos2017continuous}. So far,  the protocol has been shown to offer cloning and collision resistance, while its security against an emulation attack, in which an adversary knows the challenge-response properties of the PUK and he can access the challenges during the verification, has been analyzed in Ref. \citep{nikolopoulos2018continuous}.  The analysis of Ref. \citep{nikolopoulos2018continuous}  is rather general and does not involve any assumptions about the type of the state-discrimination strategy adopted by the adversary. Hence, the  natural question arises  whether the security bound obtained in Ref. \citep{nikolopoulos2018continuous} can be attained by means 
of  standard state-discrimination techniques, which are within reach of current or future technology. 
In the present work this question is addressed  for  unambiguous state discrimination, minimum-error discrimination and dual-homodyne detection. 

The paper is organized as follows. In Sec. \ref{sec2} we discuss briefly the EAP, while the intercept-resend attacks under consideration are formulated in  Sec. \ref{sec3}. Our main results and the robustness of the protocol are discussed in  Sec. \ref{sec4}. A summary with concluding remarks is given in Sec. \ref{sec5}.

	
\section{Authentication Scheme}
\label{sec2}

The EAP under consideration and the related verification set-up  have been discussed in detail elsewhere  \cite{nikolopoulos2017continuous,nikolopoulos2018continuous}, but for the sake of completeness we will give here a brief overview, focusing mainly on the aspects of the EAP that are directly pertinent to the present work. 
The interested reader may refer to Refs.  \cite{nikolopoulos2017continuous,nikolopoulos2018continuous}  for a detailed description.

\begin{figure*}
\centerline{\includegraphics[width=\linewidth]{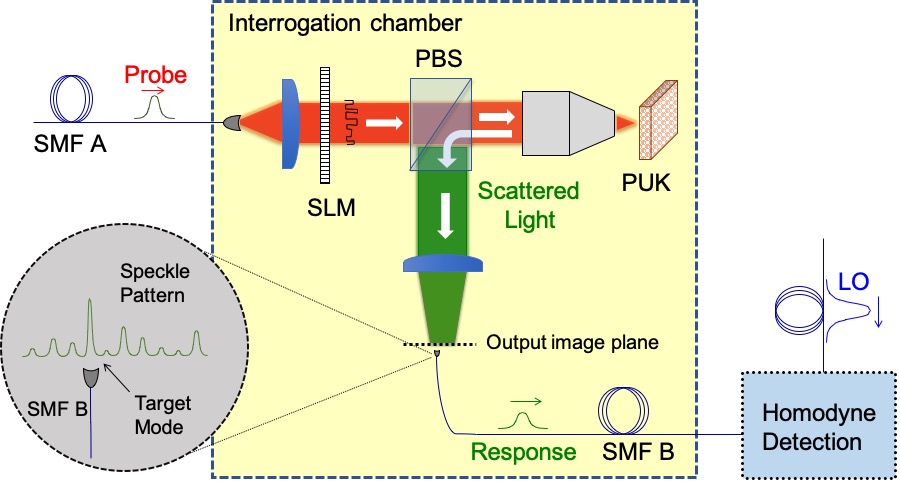}}
\caption{Schematic representation of the set-up for the EAP under consideration \cite{nikolopoulos2017continuous}. }
\label{setup:fig}
\end{figure*}

The main parts of the verification set-up are shown in Fig. \ref{setup:fig}.   A coherent  probe is directed to the interrogation chamber via a single-mode fiber (SMF A), and its wavefront is shaped by a phase-only spatial light modulator (SLM) 
thereby obtaining the  quantum challenge, which is focused on the PUK. The scattered light (speckle), is collected by means of  a polarizing beam-splitter and an objective. One of  the speckle grains is coupled to a properly positioned single-mode fiber (SMF B), which leads to a standard homodyne detection (HD) set-up. SMF B can be translated in a controlled manner at the output plane so that to collect light from different speckle grains (different target modes).  The field in the mode of the fiber is the response of the PUK to the particular challenge, and in the HD set-up, the verifier measures at random one of the two conjugate quadratures using a strong local oscillator (LO) as reference. 

The set-up operates in the diffusive limit \cite{nikolopoulos2017continuous,nikolopoulos2018continuous}, and the phase-mask of the SLM is optimized so that the intensity  of the scattered light in  SMF B is maximized. The optimization can be performed with standard algorithms \cite{Vellekoop:15, Mosk:12,Poppoff:11}, and for a fixed set-up the optimal phase mask 
${\bm \Phi}$ of the SLM, depends strongly on the PUK $\mathcal{K}$ and the chosen target mode at the output 
$s$, while it is independent of the details of the probe state \cite{defienne2014nonclassical,huisman2014controlling}. Moreover, having obtained 
the optimal phase pattern, one can apply a global phase shift to it, thereby determining the phase of the field at the target mode relative to the LO,  without affecting the intensity \cite{Huisman:15}. 
Throughout this work we assume that the probe is in a coherent state with fixed mean number of photons $\mu_{\rm P}$, which is chosen at random and independently from 
the symmetric set  
\bea
{\mathbb S}_N:=\left \{ \ket{\alpha_k}=\ket{\sqrt{\mu_\text{P}}e^{i\varphi_k}}~:~ \varphi_k := \frac{2\pi k}{N},\, k\in\mathbb{Z}_N \right \}, 
\eea
where $\mathbb{Z}_N = \{0,1,\ldots, N-1\}$ and $N\gg 1$. We associate a randomly chosen phase shift with each possible value of $k$, which is applied on the optimal phase pattern of the SLM when the probe state  $\ket{\alpha_k}$ is used. Hence, the optimal phase pattern  for the probe state  $\ket{\alpha_k}$ becomes ${\bm \Phi}_k:={\bm \Phi}+\omega_k$, where $\omega_k$ denotes the applied global shift, and is chosen at random and independently from a uniform distribution over finite set ${\bm \Omega}$. 
 In the diffusive limit, the action of the entire set-up on the input probe can 
be represented by a linear transformation, which depends on various parameters of the set-up, including the PUK,  and the (shifted) phase pattern of the SLM. The main quantity of interest is the  quantum mechanical  expectation value of the $\theta$-quadrature of the electric field in SMF B at the entrance of the HD set-up, which  is given by \cite{nikolopoulos2017continuous,nikolopoulos2018continuous}
\bea
\aver{\hat{Q}(\theta)}_k = \sqrt{2\mu_{\rm R}} \cos(\chi_k-\theta), 
\label{Q_k:eq}
\eea
where the mean number of photons $\mu_{\rm R}$ that reach the HD set-up is smaller than $\mu_{\rm P}$. 
Equation  (\ref{Q_k:eq})  encapsulates possible losses, and various other parameters pertaining to the set-up and the applied wavefront-shaping technique.
For a fixed set-up with publicly-known parameters, $\mu_{\rm R}$ can be considered fixed and known, while the phase $\chi_k$ is fully determined by the phase of the probe state $\varphi_k$, the associated random phase shift $\omega_k$, and the PUK.  For $\theta = 0$ and $\pi/2$ one obtains the expectation values of the conjugate quadratures of the field $\hat{X}$ and $\hat{Y}$, respectively.

\begin{table}[t]
\centering
\caption{Illustration of a set of CRPs used for authentication of a PUK. The set is identified by a unique identification number (see discussion in Sec. \ref{sec5}). The angles $\theta = 0$ and $\pi/2$ refer 
to the quadratures $\hat{X}$ and $\hat{Y}$ of the field in SMF B at the entrance of the HD set-up, respectively (see Fig. \ref{setup:fig}).}
\label{tab1}
\begin{tabular}{c|c|c|c}
\hline
\hline
 \multicolumn{4}{c}{{\bf Identification Number}}  \\
 \hline
 \multicolumn{2}{c|}{{\bf Challenge}}    & \multicolumn{2}{c}{{\bf Response}}             \\ 
\hline
   $k$ & Phase mask  & $\theta=0$  & $\theta = \pi/2$ \\ \hline
 0 & $ {\bm \Phi}_0$  & $\aver{\hat{X}}_0$  & $\aver{\hat{Y}}_0$\\ 
 1  &  $ {\bm \Phi}_1$ & $\aver{\hat{X}}_1$ & $\aver{\hat{Y}}_1$\\ 
 2 & $ {\bm \Phi}_2$  & $\aver{\hat{X}}_2$  & $\aver{\hat{Y}}_2$\\
 $\vdots$ &   & $\vdots$  & $\vdots$\\ 
 $N-1$ &  $ {\bm \Phi}_{N-1}$ & $\aver{\hat{X}}_{N-1}$  & $\aver{\hat{Y}}_{N-1}$\\
 \hline
 \hline
\end{tabular}
\end{table}

There are two distinct stages in the EAP under consideration i.e., the  enrollment and the verification stages (see Fig. \ref{qpuf_scheme:fig}). 
The enrollment stage takes place only once,  
before a PUK is given to a user, it is performed by the 
manufacturer,  and aims at a reliable characterization of the 
PUK with respect to its responses to a finite set of challenges. The typical list of CRPs for the protocol under consideration is given in Table \ref{tab1}.  We assume that the enroller has all the resources needed for the reliable 
estimation of the response $R_k:=\{\aver{\hat{X}}_k, \aver{\hat{Y}}_k\}$  for each challenge $C_k:=\{k, {\bm \Phi}_k\}$  \cite{nikolopoulos2017continuous}. 
After the enrollment stage, the list of CRPs is stored in a database of a server, and the PUK is given to a user. 

The verification stage takes place each time the user wishes to authenticate himself. He has to input the PUK in a potentially unsecured verification set-up, which is connected to the database over a secure and authenticated classical channel. The server sends to the verifier a sequence of $M\gg 1$ challenges, chosen at random from the available challenges, together with a sequence of $M$ angles $(\theta_1, \theta_2, \ldots, \theta_{M})$ chosen at random and independently  from  a uniform distribution over $\{0,\pi/2\}$. The PUK is interrogated sequentially by the $M$ challenges. In the $j$th challenge $C_{k_j}:=\{k_j, {\bm \Phi}_{k_j}\}$,  
the verifier measures the $\theta_j-$quadrature of the field in SMF B.  
Assuming very strong LO field, the outcome of the measurement in the HD set-up follows a normal distribution $\mathcal{N}(\braket{\hat{Q}(\theta_j)}_{k_j}, \sigma^2)$, which is centered at the expectation value of the measured quadrature  and its standard deviation $\sigma=1/\sqrt{2\eta}$, is determined by the detection efficiency $\eta$ of the HD set-up. 
The  outcomes from all of the $M$ measurements are returned to the server, over the classical channel. Acceptance or rejection of the PUK is decided upon the fraction of outcomes $p_\text{in}$ that 
fall within an interval (bin) of width $\Delta$, which is centered at the expectation value of the measured quadrature $\aver{\hat{Q}(\theta_j)}_{k_j}$.  The theoretically expected probability for an outcome to fall inside the interval is given by \cite{nikolopoulos2018continuous}
\begin{align}
P_\text{in}^{(0)}=\text{erf}\left(\frac{\bar{\Delta}}{2\sqrt{2}}\right) 
\label{P0in}
\end{align}
with $ \bar{\Delta}=\Delta/\sigma$, which is independent of $k$ and $\theta$. 
For sufficiently large values of $M$, one can ensure with high confidence, that for the true PUK and in the absence of cheating, the empirical probability 
$p_\text{in}$ will lie in an interval of size $2\varepsilon$ around the theoretically expected probability 
$P_\text{in}^{(0)}$ \cite{nikolopoulos2017continuous,nikolopoulos2018continuous}. Hence, the PUK is accepted if $|p_\text{in}-P_\text{in}^{(0)}|<\varepsilon$, and is rejected otherwise. 
As we will see later on, the security parameter $\varepsilon$ plays a central role in the protocol, as it determines 
the regime of  parameters for which an attack can be detected by the verifier. 


\section{Intercept-Resend Emulation Attacks}
\label{sec3}

\begin{figure*}
\centerline{\includegraphics[width=0.8\linewidth]{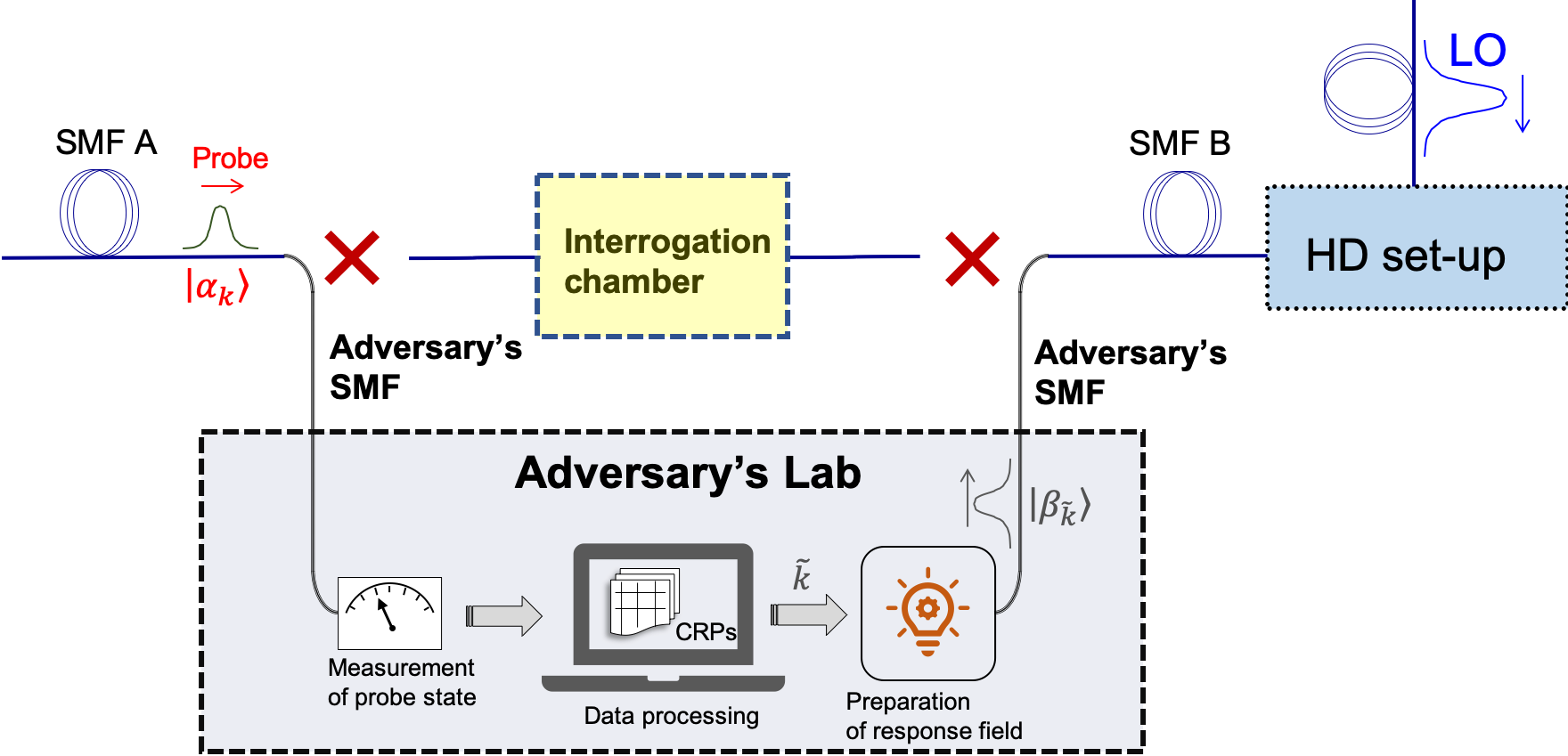}}
\caption{Schematic representation of the attack under consideration \cite{nikolopoulos2018continuous}. The adversary  has a copy of the set of the numerical CRPs, from which the challenges are chosen at random. 
He intercepts each one of the $M$ incoming probe states, and measures it in order to deduce the value of $k$ (see table \ref{tab1}). Based on the outcome of his measurement the adversary makes an educated guess about $k$, say $\widetilde{k}$, and he looks $\widetilde{k}$ up in the set of CRPs in order to find the corresponding expected response $R_{\widetilde{k}}$. Finally, the adversary prepares and sends to the HD set-up of the verifier a quantum state that will induce statistics consistent with $R_{\widetilde{k}}$.}
\label{attack_scheme:fig}
\end{figure*}

The emulation attacks under consideration follow closely the general theoretical  framework of Ref. \cite{nikolopoulos2018continuous}, which is summarized in Fig. \ref{attack_scheme:fig}. We assume that the adversary does not have access to the preparation of the probe states, the actual PUK, the SLM, or the HD set-up. However, he has obtained somehow a copy of all the possible CRPs to be used in the authentication of the PUK, and moreover he has obtained access to the incoming and outgoing SMFs, without being detected. Hence, if the adversary were able to estimate exactly the probe state, he would be able to 
impersonate successfully the holder of the PUK, by looking at the CRPs and by sending to the HD set-up the expected response state. However, quantum physics does not allow for perfect discrimination between non-orthogonal 
quantum states, or the measurement of non-communting quantum observables  with arbitrary accuracy. As a result, it is inevitable for the adversary's intervention to introduce errors in the HD performed by the verifier.

Given that the probe states in the $M$ queries are chosen at random and independently from a uniform 
distribution over the set ${\mathbb S}_N$,  from now on we can focus on one of the queries and let $\ket{\alpha_k}$ denote the incoming probe state. The adversary's task is to obtain an estimate 
 of the integer $k$, which determines fully the challenge and the expected response. To this end, he performs a measurement on the incoming probe state,  and based on the outcome, the adversary makes an educated guess about $k$, to be denoted by $\widetilde{k}$. For the reasons discussed above, this guess may or may not be equal to the actual $k$ used by the verifier, and thus the adversary's  intervention will affect the probability with which the verifier obtains an outcome in the expected bin. 
 More precisely, instead of Eq. (\ref{P0in}) one has \cite{nikolopoulos2018continuous}
 \begin{subequations}
 \label{PVerIn}
 \begin{widetext}
 \bea
P(\text{in}|k,\widetilde{k},\theta)=\frac{1}{2} 
\left\{ \text{erf}\left [
\frac{2\left (\aver{\hat{Q}(\theta)}_k-\aver{\hat{Q}(\theta)}_{\widetilde{k}}\right )+\Delta}{2\sqrt{2}\sigma}
\right ] 
-\text{erf}\left [
\frac{2\left (\aver{\hat{Q}(\theta)}_{k}-\aver{\hat{Q}(\theta)}_{\widetilde{k}} \right )-\Delta}{2\sqrt{2}\sigma}\right] \right \},
\label{PVerIn1}
\eea
\end{widetext}
where $\hat{Q}_{k}(\theta)$ and $\hat{Q}_{\widetilde{k}}(\theta)$ denote the $\theta-$quadratures of the response field, when the adversary deduces the correct and the wrong value of $k$. Both of them are given by Eq. (\ref{Q_k:eq}) for $k$ and $\widetilde{k}$, respectively. 
 
 Given that the verifier samples from both quadratures at random and independently, after $M$ queries the verifier will obtain an estimate of the average  probability  \cite{nikolopoulos2018continuous}
 \begin{widetext}
\bea
P_{\rm in} &=& \sum_{k=0}^{N-1} \sum_{\widetilde{k}=0}^{N-1} \sum_{\theta\in\{0,\pi/2\}} P({\rm in},k,\widetilde{k},\theta) 
= 
(1-P_{\rm err} )  P_{\rm in}^{(0)} +\frac{1}{2N} \sum_{k}\sum_{\widetilde{k}\neq k}
P(\widetilde{k}|k)\sum_{\theta=0,\pi/2} 
P({\rm in}|k,\widetilde{k},\theta), 
\label{PVerIn2}
\eea 
\end{widetext}
\end{subequations}
where $P({\rm in}|k,\widetilde{k},\theta)$ is given by Eq. (\ref{PVerIn1}) and 
\bea
P_{\rm err}:=\sum_k P(k,\widetilde{k} \neq k)  
=1-\frac{1}{N} \sum_k P(\widetilde{k} = k|k)
\eea
is the  probability for the adversary to make an error in his estimation of $k$.

\begin{figure}[t]
		\centering
		\includegraphics[width=0.49\textwidth]{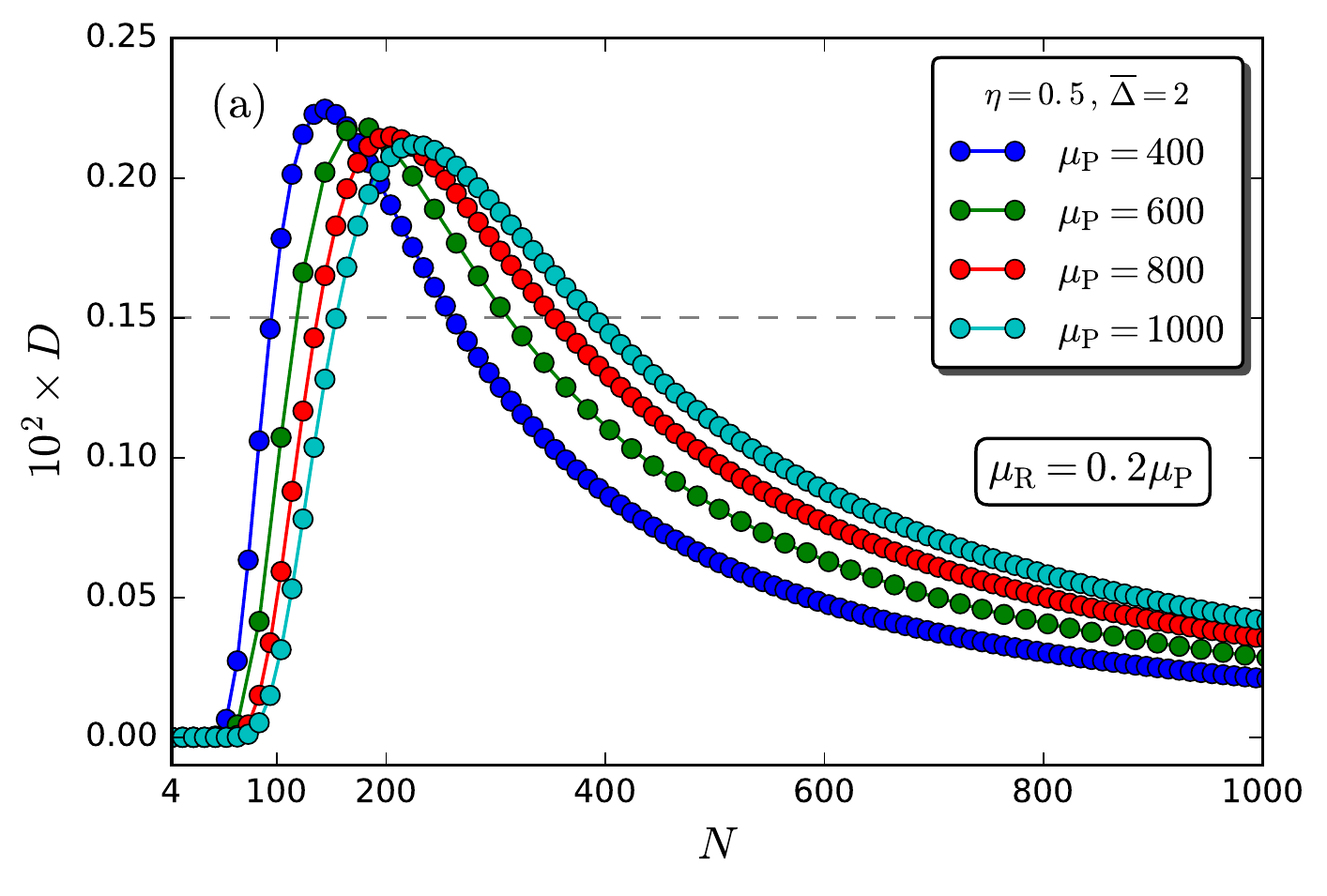}
		\includegraphics[width=0.49\textwidth]{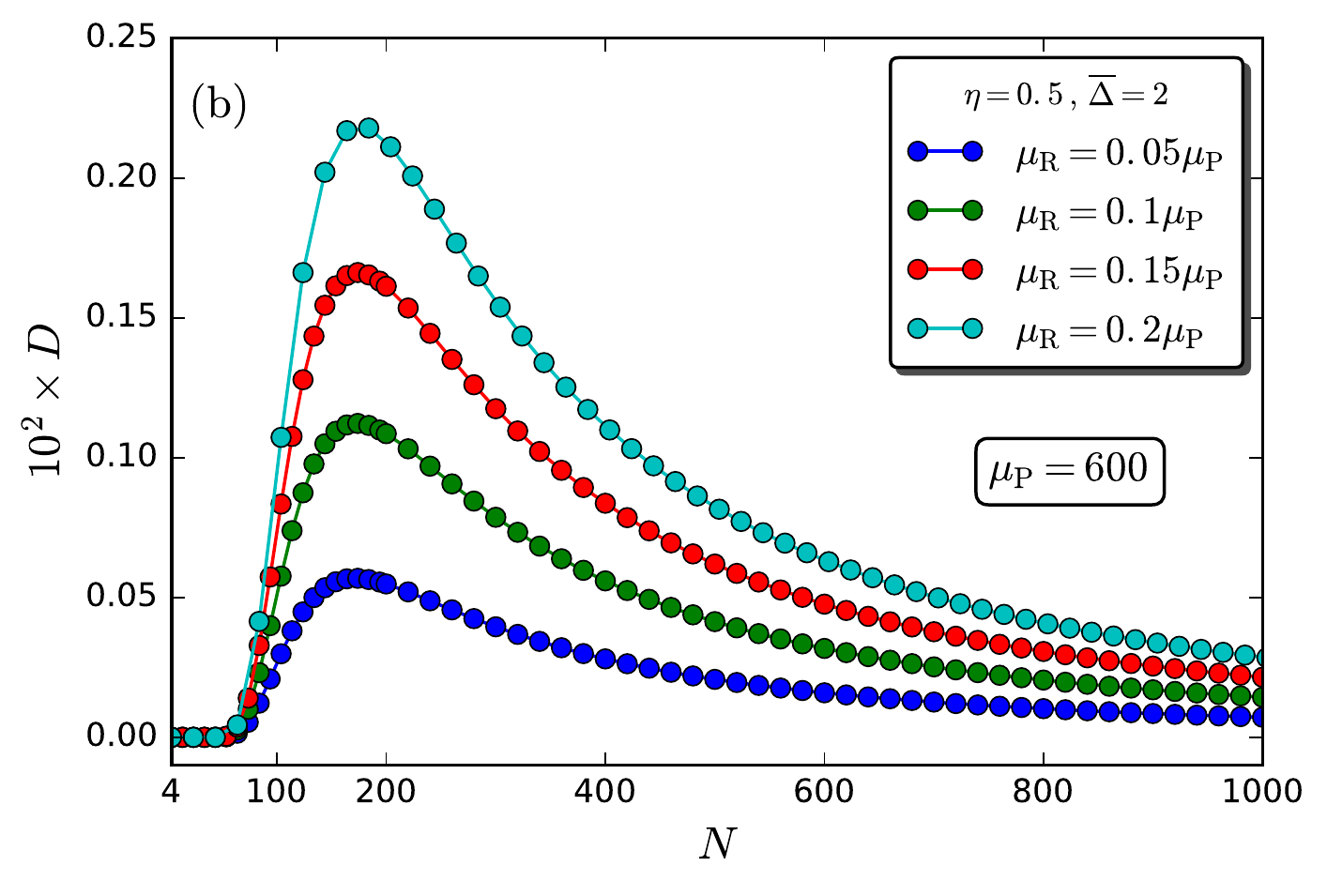}
		\caption{ The lower bound $D_{\rm low}$, as a function of the number of different probe states $N$, for various values of $\mu_P$ but fixed ratio 
$\mu_{\rm R}/\mu_{\rm P}$ (a); and various values of $\mu_R$ for fixed $\mu_{\rm P}$ (b). The vertical dashed line marks the security threshold $2\varepsilon = 15\times 10^{-4}$, and the protocol is secure against any intercept-resend attack for $D_{\rm low} > 2\varepsilon$. }
		\label{DLow:fig}
	\end{figure}

The probability $P_{\rm in}$ takes into account all of the events irrespective of whether the adversary deduced the correct value of $k$ or not. In the second equality of Eq. (\ref{PVerIn2}) we have taken into account the uniform distribution of $k$ over ${\mathbb Z}_N$ and of 
$\theta$ over $\{0,\pi/2\}$. Moreover, we have used the fact that the state of the probe and the outcome of the adversary are independent of the quadrature to be measured by the verifier.    
A straightforward calculation \cite{nikolopoulos2018continuous} shows that  $P_{\rm in}$  is always smaller than the corresponding expression in the absence of cheating [which is given by Eq. (\ref{P0in})], and the  difference 
\begin{subequations}
\label{all_sec_cond}
\be
D:=P_{\rm in}^{(0)}-P_{\rm in},
\label{D:def}
\ee
basically quantifies the adversary's intervention. The adversary's intervention will be detected by the verifier 
when the observed deviations from $P_{\rm in}^{(0)}$ exceed the statistical deviations i.e., when 
\bea
D>2\varepsilon. 
\label{sec_cond}
\eea
\end{subequations}
As mentioned above, the security threshold $2\varepsilon$ is determined by the sample size $M$ in the verification stage while, in general,  the  
exact values of $P_{\rm err}$, $P_{\rm in}$, and $D$, depend on the measurement applied by the adversary.  
However, working along the lines of Ref.  \cite{nikolopoulos2018continuous} one  readily obtains a lower bound on $P_{\rm err}$, which in turn yields the following lower bound on $D$ 
\bea
D\geq P^{{\rm (low)}}_{\rm err}
\left [
P^0_\text{in}-P_{\max }({\rm in}|{\rm error})
\right ]
:=D_{\rm low}, 
\label{LowBuneq}
\eea
where  
$P_{\max}({\rm in}|{\rm error}) := \max_{k,\widetilde{k}}
\left \{ 
P(\text{in}|k,\widetilde{k}) 
\right \}_{\widetilde{k}\neq k} $.
This bound is expected to hold irrespective of the adversary's measurement, and its  
dependence on various parameters of the protocol has been discussed in detail elsewhere \cite{nikolopoulos2018continuous}. 
In particular, it has been shown that for any given combination of parameters $\{\varepsilon,  \Delta, \mu_P, \mu_R\}$, the secure operation region for the protocol becomes wider for increasing $\eta$. Typical HD set-ups have $\eta>0.5$, and  $\eta=0.5$ can be considered as the worst-case scenario for the security of the protocol. The size of the bin width $\Delta$ relative to the detection efficiency is a free parameter that can be chosen at will, so that to optimize the performance of the protocol. For any given combination of  $\{\varepsilon, \eta,  \mu_P, \mu_R\}$ it has been shown that the optimal value  is $\Delta\simeq 2\sigma$. Hence, although our simulations have been extended to various combinations of $\eta$ and $\Delta$, throughout the present work we focus on  numerical results for $\eta = 0.5$ and $\Delta = 2\sigma$. To facilitate the analysis of our results for the three different attacks, in Fig. \ref{DLow:fig} we plot the lower bound $D_{\rm low}$ as a function of $N$, for different values $\mu_{\rm P}$ and $\mu_{\rm R}$. As explained in Ref. \cite{nikolopoulos2018continuous}, the depicted asymmetric bell-shape stems from the fact that  $D_{\rm low}$ 
 is a product of two functions of $N$ with opposite monotonicity [see Eq. (\ref{LowBuneq})]. More precisely, $P^{{\rm (low)}}_{\rm err}$ increases with increasing $N$ approaching a non-zero value determined by the classical limit, while $P_{\rm in}^{(0)}-P_{\max }({\rm in}|{\rm error})$ decreases with increasing $N$, approaching zero for large values of $N$.
A crucial parameter for the security of the protocol is the mean number of photons that reach the HD set-up. The larger  $\mu_{\rm R}$ is for a fixed $N$,  the easier the verifier can discriminate between $\aver{\hat{Q}_k(\theta)}$ and $\aver{\hat{Q}_{\widetilde{k}\neq k}(\theta)}$.  
In general, $\mu_{\rm R}$ can be  increased in two different ways: (i) increasing 
$\mu_{\rm P}$ for fixed losses in the set-up; (ii) improving the set-up, thereby  reducing the losses. Figures  \ref{DLow:fig}(a) and  \ref{DLow:fig}(b), refer to these two cases, and in both of them we see that for a given $\varepsilon$ the secure region gets wider with increasing $\mu_R$, which is in agreement with 
the aforementioned role of $\mu_{\rm R}$ in the discrimination of different responses.

\subsection{Dual Homodyne-detection Attack}
\label{sec3a}
In a dual homodyne-detection attack (DHA), for each probe the adversary performs a joint measurement of the non-commuting observables $\{\hat{X}_k,\hat{Y}_k\}$, by means of an eight-port interferometric set-up  \cite{leonhardt1995uncertainty, schleich2011quantum}, thereby obtaining 
a bivariate random variable $(x,y)$, which follows the two-dimensional normal 
distribution
\begin{subequations}
\label{Alljoint_prob:eq}
\bea
P^\text{(DH)}_k(x,y)=\frac{1}{2\pi} \exp\left(-\frac{\left[x-\aver{\hat{X}_A}_k\right]^2+\left[y-\aver{\hat{Y}_A}_k\right]^2}{2}\right)\nonumber\\
\label{joint_prob:eq}
\eea
where $\aver{\hat{X}_A}_k$ and $\aver{\hat{Y}_A}_k$ refer to the expectation values of the quadratures of the electric field in SMF A, when the field is in state $\ket{\alpha_k}$. Assuming negligible losses in SMF A and in the  set-up of the adversary we have 
\bea
&&\aver{\hat{X}_A}_k = \bra{\alpha_k} \hat{X}_A \ket{\alpha_k} = \sqrt{2\mu_{\rm P}} \cos(\varphi_k),\nonumber \\
&&\aver{\hat{Y}_A}_k = \bra{\alpha_k} \hat{Y}_A \ket{\alpha_k} = \sqrt{2\mu_{\rm P}} \sin(\varphi_k). 
\label{jXaYa:eq}
\eea
\end{subequations}
Moreover, in Eq. (\ref{joint_prob:eq}) we have assumed perfect detection efficiency for the adversary $(\sigma_{\rm DH} = 1)$, which is the worst-case scenario for the security of the protocol. Hence,  when the efficiency in the HD set-up of the verifier 
is $\eta\geq 0.5$ we have $\sigma\leq \sigma_{\rm DH}$. This suggests that, 
for either of the quadratures, the adversary samples from a distribution, which is at least as broad as the distribution sampled by the verifier.  

\begin{figure}
\centerline{\includegraphics[width=0.9\linewidth]{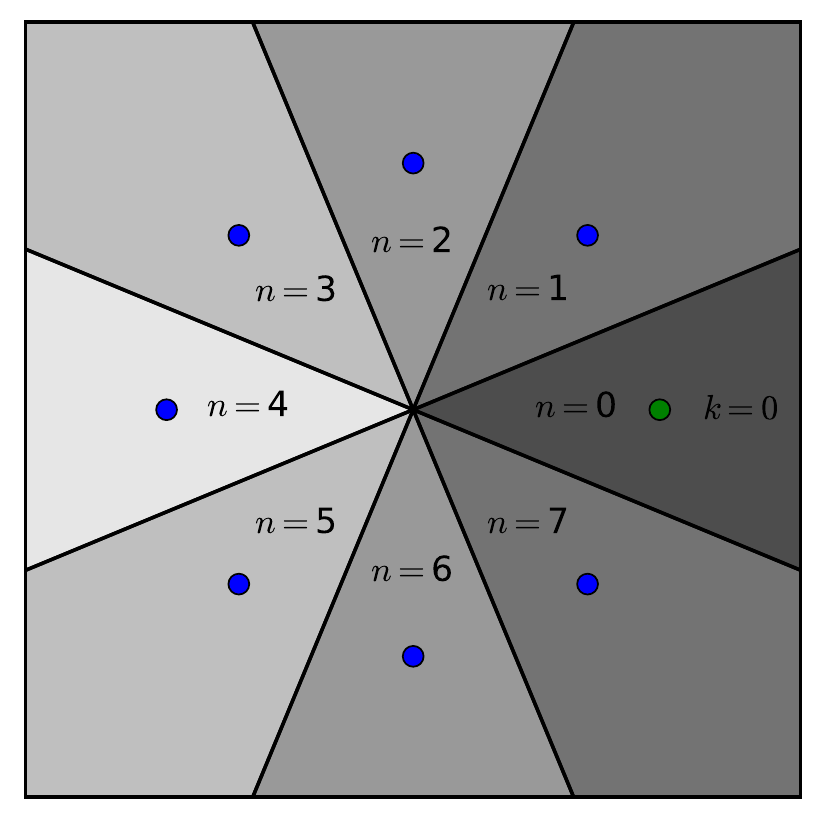}}
\caption{Discretization of the phase space for $N=8$. The segment for the actual probe state $k=0$ (green) and all other possible probe states (blue) are shown. The coloring displays the probabilities $P^\text{DH}(\widetilde{k}|k=0)$  for $\widetilde{k}=0,1,\ldots,N-1$ (dark: high probability).}
\label{fig_disc1}
\end{figure}
	
To proceed further, we have to define a strategy for the adversary to make an educated guess  for the probe stare $\ket{\alpha_k}$, given the random outcome of the dual-homodyne detection $(x,y)$. Recall here that  the adversary has obtained a copy of the list of the CRPs for the particular PUK, which allows him to discretize the phase space as shown in Fig. \ref{fig_disc1} . The $k$-th sector corresponds to the probe state  $\ket{\alpha_k}$. It is centered at 
$\{\aver{\hat{X}_A}_k,\aver{\hat{Y}_A}_k\}$, and its angular extension is from $\varphi_{k}-\pi/N$  to 
$\varphi_{k}+\pi/N$, while its radius $\rho$ extends from 0 to $\pm \infty$. So, when the random outcome 
$(x,y)$ falls in the $\widetilde{k}$-th sector, the adversary concludes that the probe state was $\ket{\alpha_{\widetilde{k}}}$, and thus the response expected by the verifier is $R_{\widetilde{k}}$. 
Working in polar coordinates, the probability for the random outcome 
$(x,y)$ to fall within the $\widetilde{k}$-th sector given that the probe state was 
$\ket{\alpha_k}$ is given by the integral 
\begin{widetext}
\bea
	P^\text{(DH)}(\widetilde{k}|k)= \frac{1}{2\pi}  \int_{\varphi_{\widetilde{k}}-\frac{\pi}{N}}^{\varphi_{\widetilde{k}}+\frac{\pi}{N}}\text{d}\gamma\int_{0}^{\infty} \rho \exp\left [-\frac{\rho^2+2\mu_{\rm P}-\sqrt{8\mu_{\rm P}}\rho\cos(\gamma-\varphi_{k})}{2}\right ]\text{d}\rho, 
	\label{P_DH1:eq}
\eea
\end{widetext}
where we have set $x=\rho\cos(\gamma)$, $y=\rho\sin(\gamma)$, and used 
Eqs. (\ref{Alljoint_prob:eq}). 
Setting $\gamma^\prime=\gamma-\varphi_k$ one can readily confirm that this  probability depends only on the difference 
$ n := (\tilde{k} - k)\mod N$. Therefore, the probability for the adversary to deduce the wrong probe state is given by 
	\bea
	P^\text{(DH)}_\text{err}=1-\sum_{k=0}^{N-1} p(k) P^\text{DH}(\widetilde{k} = k|k)=1 -P^\text{DH}(\widetilde{k} = k|k).	\nonumber\\
	\label{PDHerr}
\eea

After the measurement, the adversary prepares and sends to the verifier a coherent state 
$\ket{\beta_{\widetilde{k}}}$, which is consistent with his educated guess 
about the probe state, and will induce the expected statistics at the HD set-up of the verifier. The average probability $P^\text{(DH)}_\text{in}$ for the state $\ket{\beta_{\widetilde{k}}}$ to result in an outcome within the expected bin is given  by Eq. (\ref{PVerIn2}),  
after substituting  $P(\widetilde{k}|k)$ by $P^{\rm (DH)}(\widetilde{k}|k)$, which is given by Eq. (\ref{P_DH1:eq}). Hence, the adversary's intervention will introduce errors in the 
estimate to be obtained by the verifier, which are quantified by the difference 
\begin{widetext}
\bea
D^{\rm (DH)}:=P^{(0)}_\text{in}-P^\text{(DH)}_\text{in}=P_{\rm err}^{\rm (DH)}  \left \{ P_{\rm in}^{(0)} - 
 \frac{1}{P_{\rm err}^{\rm (DH)}} \left [ \frac{1}{2N} \sum_{k}\sum_{\widetilde{k}\neq k}
P^{\rm (DH)}(\widetilde{k}|k)\sum_{\theta=0,\pi/2} 
P({\rm in}|k,\widetilde{k},\theta)\right ]\right \}.
\label{DDH}
\eea
\end{widetext}
	
The term inside the brackets is the joint probability for the adversary to deduce the wrong value for $k$, and the outcome of the HD of the verifier to fall within the expected bin. 


\subsection{Unambiguous State-Discrimination Attack}	
\label{sec3b}

Assume now that the adversary applies an ideal  unambiguous discrimination (UD) measurement to each probe, 
in order to deduce which  of the $N$ possible coherent states is used by the verifier \cite{chefles1998optimum,ivanovic1987differentiate,peres1988differentiate,dieks1988overlap}. 
The adversary will either obtain a conclusive result, or an inconclusive result. In the former case, he learns the actual state of the probe with certainty, whereas in the latter he learns nothing about the incoming state, which is  destroyed by the measurement, and additional measurement on it will not provide any useful  information. Given that the adversary has the list of CRPs,  a conclusive result implies that the adversary knows precisely the response expected by the verifier, whereas in the case of an inconclusive result,  all of the possible responses are equally probable. 

The minimum probability for an inconclusive result is \cite{chefles1998optimum}
\bea
P_{\rm inc}=1- N\cdot \text{min}_{r\in{\mathbb Z}_N}\left(\frac{1}{N}\sum_{j=0}^{N-1}e^{-{\rm i}2\pi jr/N}\cdot e^{\vert\alpha\vert^2(e^{{\rm i}2\pi j/N}-1)}\right), 
\label{Pinc}
\eea
which is independent of the value of $k$ i.e., independent of the probe state. 
In the case of a conclusive outcome  the adversary will prepare and send a coherent state, which will induce the  expected statistics at the HD set-up of  the verifier. Hence, the probability for the verifier to get a result inside the bin, given that the adversary obtained a conclusive outcome, is equal to $P_{\rm in}^{(0)}$ [see Eq. (\ref{P0in})].  On the contrary, in the case of 
an  inconclusive result, the adversary has no information on  the probe state or on the response  expected by the verifier. Given that the verifier expects a response state for each probe, the adversary has to send a state to the HD set-up of the verifier. Hence, we assume that in the case of an inconclusive outcome, the adversary has to choose at random one of the $N$ equally probable responses, and send a coherent state $\ket{\beta_{\widetilde{k}}}$ that will induce the corresponding statistics at the HD set-up of the verifier.   This choice is independent of the probe state, and thus the conditional probability for the random choice to yield $\widetilde{k}$, given an inconclusive result on the input state $\ket{\alpha_k}$ is given by 
\bea
P^{\rm (UD)}({\widetilde k}|{\rm inc},k)=P^{\rm (UD)}({\widetilde k}) = \frac{1}N. 
\label{P_condUD:eq}
\eea 
Using this equation, the overall error probability reads 
\bea
P_{\rm err}^{\rm (UD)} 
= P_{\rm inc}\frac{N-1}{N}.
\label{PUDerr}
\eea
Using the fact that a conclusive outcome always yields the correct value of $k$, while the occurrence  of a conclusive or inconclusive result is independent of $k$, one readily obtains 
\bea
P^{\rm (UD)}({\widetilde k}\neq k|k) 
= \frac{P_{\rm inc}}N.
\label{P_condUD:eq}
\eea
 
The average probability for the verifier to get an outcome inside the expected bin after an unambiguous-state-discrimination attack (UDA) is 
given by Eqs. (\ref{PVerIn}), after replacing $P({\widetilde k}|k)$ by  Eq. (\ref{P_condUD:eq}) and $P_{\rm err}$ by $P^\text{(UD)}_\text{err}$. 
Accordingly, the deviations  that will be observed by the verifier are quantified by the difference 
\begin{widetext}
\bea
D^{\rm (UD)}=P_{\rm err}^{\rm (UD)}  \left \{ P_{\rm in}^{(0)} - 
 \frac{1}{P_{\rm err}^{\rm (UD)}} \left [ \frac{1}{2N} \sum_{k}\sum_{\widetilde{k}\neq k} P^{\rm (UD)}({\widetilde k}|k)
\sum_{\theta=0,\pi/2} 
P({\rm in}|k,\widetilde{k},\theta)\right ]\right \}.
\label{DUD}
\eea
\end{widetext}
where $P({\rm in}|k,\widetilde{k},\theta)$ is given by Eq. (\ref{PVerIn1}). 


\subsection{Minimum-error State Discrimination Attack}	
\label{sec3c}

The third attack we consider relies on a minimum-error state-discrimination measurement  \cite{barnett2009quantum,weir2018optimal}. 
Each of the possible outcomes ${\widetilde k}\in\{0, 1, \ldots N-1\}$ , is characterized by a corresponding probability operator measure element $\hat{\Pi}_{\widetilde{k}}$, also referred to as a positive operator-valued measure.  
The probability for the adversary to obtain outcome $\widetilde{k}$, given that the probe state $\ket{\alpha_k}$ was used,  is 
\bea
P^\text{(SR)}(\widetilde{k}|k) 
= 
\text{Tr}\left(\hat{\rho}_k\widehat{\Pi}_{\widetilde{k}}\right),
\label{P_SR1:eq}
\eea
where  $\hat{\rho}_k:=\ket{\alpha_k}\bra{\alpha_k}$.
When the adversary obtains the outcome $\widetilde{k}$, he concludes that the probe  was also in state 
$\ket{\alpha_{\widetilde{k}}}$, and thus  the response expected by the verifier is 
$R_{\widetilde{k}}=\{\aver{\hat{X}_{\widetilde{k}}}, \aver{\hat{Y}_{\widetilde{k}}}\}$.  

Given that the authentication scheme works with the symmetric set of states ${\mathbb S}_N$, the optimal measurement that 
minimizes the error probability is the square-root measurement \cite{barnett2009quantum,weir2018optimal}, where the operator 
$\widehat{\Pi}_k$ is given by 
	\begin{align}
		\widehat{\Pi}_k:=\frac{1}{N}\hat{\rho}^{-\frac{1}{2}}\hat{\rho}_k\hat{\rho}^{-\frac{1}{2}} \, ,
		\label{Pi_k:eq}
	\end{align}
with $\hat{\rho} := \sum_k \hat{\rho}_k/N$. 	
The average probability for the adversary to deduce the wrong probe state  is given by
\bea
P^\text{(SR)}_\text{err}=\frac{1}{N}\sum_{k=0}^{N-1} \sum_{\widetilde{k}
\neq k}P^\text{(SR)}(\widetilde{k}|k)  =   1-\frac{1}{N}\sum_{k=0}^{N-1}\text{Tr}\left(\hat{\rho}_k\hat{\Pi}_k\right).\nonumber\\
\label{PSRerr}
\eea

As in the previous attacks, given the outcome $\widetilde{k}$, 
the adversary prepares and sends to the verifier a coherent state that 
will induce statistics compatible with the response  $R_{\widetilde{k}}$. The probability $P^\text{(SR)}_\text{in}$ 
for the verifier to obtain an outcome within the expected bin is given  by Eqs. (\ref{PVerIn}),  
after replacing  $P(\widetilde{k}|k)$ by $P^{\rm (SR)}(\widetilde{k}|k)$, which is given by Eqs. (\ref{P_SR1:eq}) and (\ref{Pi_k:eq}). Hence, the adversary's intervention will introduce errors in the 
estimate to be obtained by the verifier, which are quantified by the difference 
$D^\text{(SR)}:=P^{(0)}_\text{in}-P^\text{(SR)}_\text{in}$. Using Eq. (\ref{PVerIn2}) we have for the square-root-measurement attack (SRA)
\begin{widetext}
\bea
D^{\rm (SR)}=P_{\rm err}^{\rm (SR)}  \left \{ P_{\rm in}^{(0)} - 
 \frac{1}{P_{\rm err}^{\rm (SR)}} \left [ \frac{1}{2N} \sum_{k}\sum_{\widetilde{k}\neq k}
P^{\rm (SR)}(\widetilde{k}|k)\sum_{\theta=0,\pi/2} 
P({\rm in}|k,\widetilde{k},\theta)\right ]\right \}.
\label{DSR}
\eea
\end{widetext}


\section{Numerical results and discussion}	
\label{sec4}

We have performed extensive simulations on the aforementioned three different types of intercept-resend attacks, and in this section we discuss our main findings on the security of the proposed protocol. As discussed in Sec. \ref{sec3}, the protocol is secure when the difference of probabilities (\ref{D:def}) exceeds 
the security threshold $2\varepsilon$, attained by sampling.  Hence, for a given value of $\varepsilon$, the main quantity of interest for the security of the protocol is the difference of probabilities $D$, which for the three different attacks is given in Eqs. (\ref{DDH}), (\ref{DUD}) and (\ref{DSR}).
A lower bound on $D$ has been obtained in Ref. \cite{nikolopoulos2018continuous}, through the Holevo bound and Fano's inequality, and it is given in Eq. (\ref{LowBuneq}).  In all of the cases, we see that $D$ is of the form 
$D=P_{\rm err}[P_{\rm in}^{(0)} - P({\rm in}|{\rm error})]$, where $P_{\rm err}$ is the error probability in each case, 
$P_{\rm in}^{(0)}$ is the probability for the verifier to obtain an outcome inside the bin in the absence of any attack [see Eq. (\ref{P0in})], and  $P({\rm in}|{\rm error})$ is the corresponding probability in the presence of the attack, and given that the adversary has made in error in deducing the right probe state.

\begin{figure*}
\centerline{\includegraphics[width=0.5\linewidth]{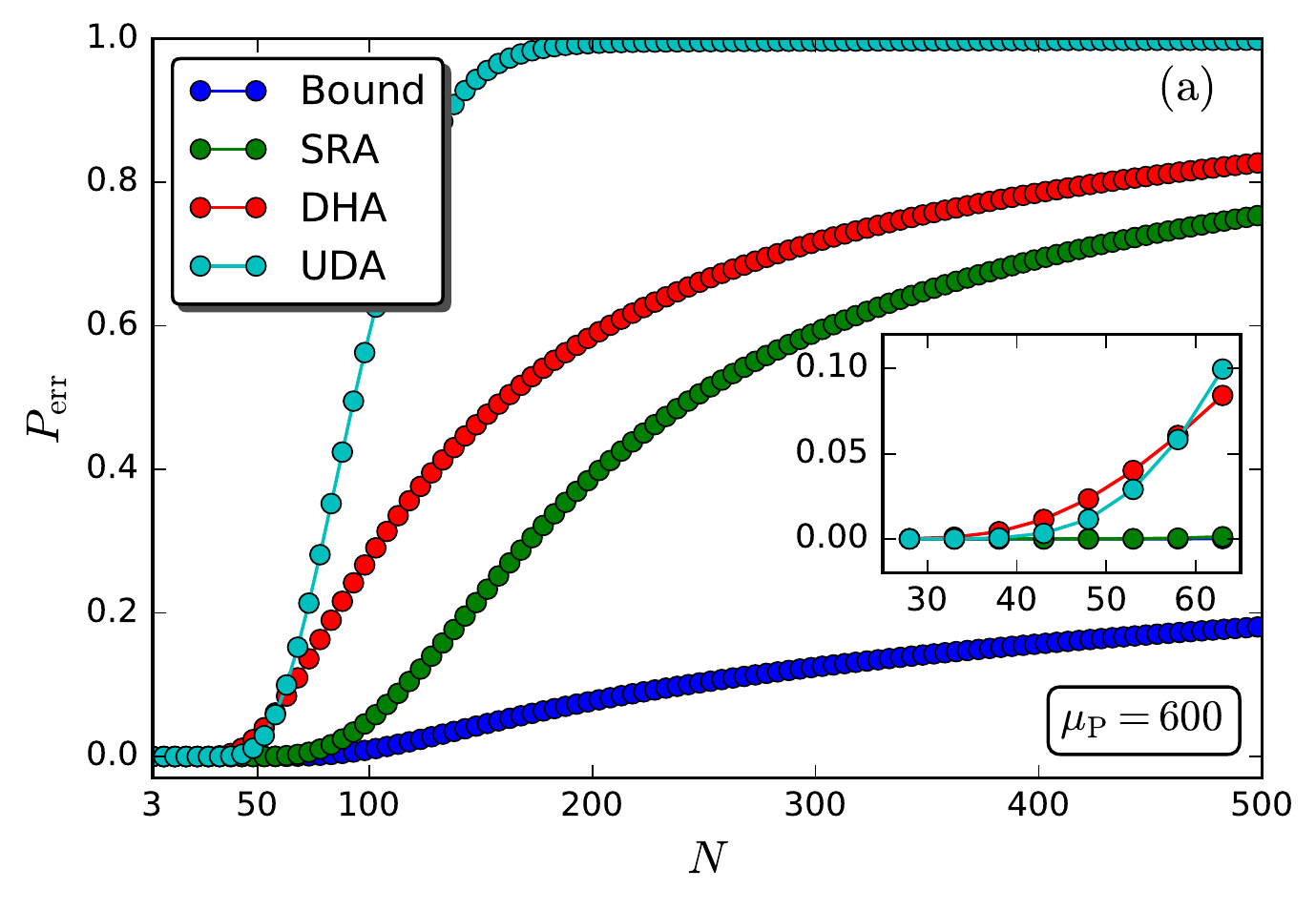}
\includegraphics[width=0.5\linewidth]{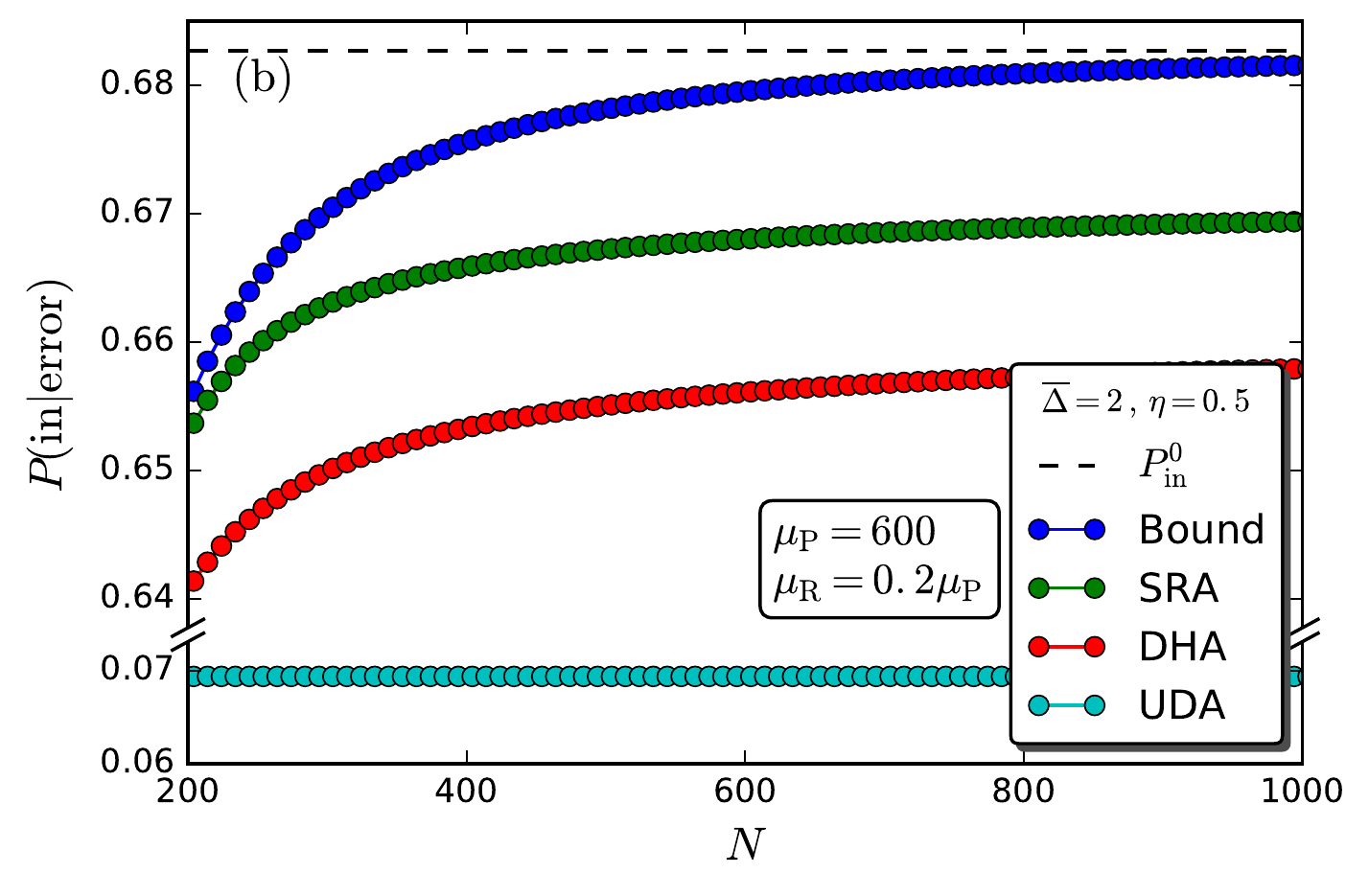}}
\caption{(a) Probability for the adversary to deduce the wrong probe state in the case of  
dual-homodyne-detection attack (red curve, DHA), square-root-measurement attack (green curve, SRA) and unambiguous-state-discrimination attack (gray curve, UDA), as a function of the number of different probe states $N$. The lower bound is also shown (blue curve). (b) Conditional probability $P(\text{in}|\text{error})$ for the verifier to obtain a result inside the expected bin given that the adversary inferred the wrong probe state, as function of $N$ and for different attacks. The maximum value $P_{\max}(\text{in}|\text{error})$ is also shown (blue curve), and all of the probabilities are compared to the probability $P^0_\text{in}$ in the absence of an attack (dashed vertical line).}
\label{Perr:fig}
	\end{figure*}

Both of $P_{\rm err}$ and $P({\rm in}|{\rm error})$ depend on the attack under consideration. Hence, it is instructive to look at them separately,  in comparison with  the corresponding quantities determining the lower bound in Eq.  (\ref{LowBuneq}), before we discuss the dependence of $D$ on the various parameters of the protocol.  
In Fig. \ref{Perr:fig}(a) we plot the probability for the adversary to deduce the wrong probe state, as a function of the number of different probe states $N$ used in the protocol, and for fixed mean number of photons in the probe. The error probabilities for the three attacks are given by Eqs. (\ref{PDHerr}), (\ref{PUDerr}), and (\ref{PSRerr}), 
while for the sake of comparison, we also show the lower bound on the error probability $P_{\rm err}^{\rm (low)}$ (blue curve). In all of the cases we find a monotonic increase of the error probability with $N$.    	
For $N < 50$, the error probabilities for all of the three attacks are very close to the lower bound, and they start deviating from it as we increase $N$. The deviation is slower in the case of the square-root-measurement attack (green curve), which is practically indistinguishable from the lower bound (blue curve) up to $N\simeq 60$, and it remains close to it for values of $N$ up to about $100$. The error probability for the unambiguous-state-discrimination attack  exhibits the fastest increase, and $P_{\rm err}^{\rm (UD)}\simeq 1$ for $N\simeq 100$ (gray curve). The dual-homodyne-detection attack stands between  the other two attacks. The same behavior has been found for other values of $\mu_{\rm P}$, and the main difference is that the error probability increases slower with $N$, for increasing values of $\mu_{\rm P}$.

Let us turn now to the other quantity of interest, namely the conditional
probability $P({\rm in}|{\rm error})$ relative to $P_{\rm in}^{(0)}$. 
As depicted in Fig. \ref{Perr:fig}(b),  for either of the three attacks under consideration  $P({\rm in}|{\rm error})$ increases with $N$, approaching an asymptotic value which is below $P_{\rm in}^{(0)}$ as well as below the asymptotic value of $P_{\rm err}$ for the same attack [compare to Fig. \ref{Perr:fig}(a)]. This is in contrast to the behavior of 
$P_{\max}({\rm in}|{\rm error})$ (blue curve), which approaches $P_{\rm in}^{(0)}$ (dashed vertical line) as we 
increase $N$.

\begin{figure*}
		\centering
		\includegraphics[width=0.49\textwidth]{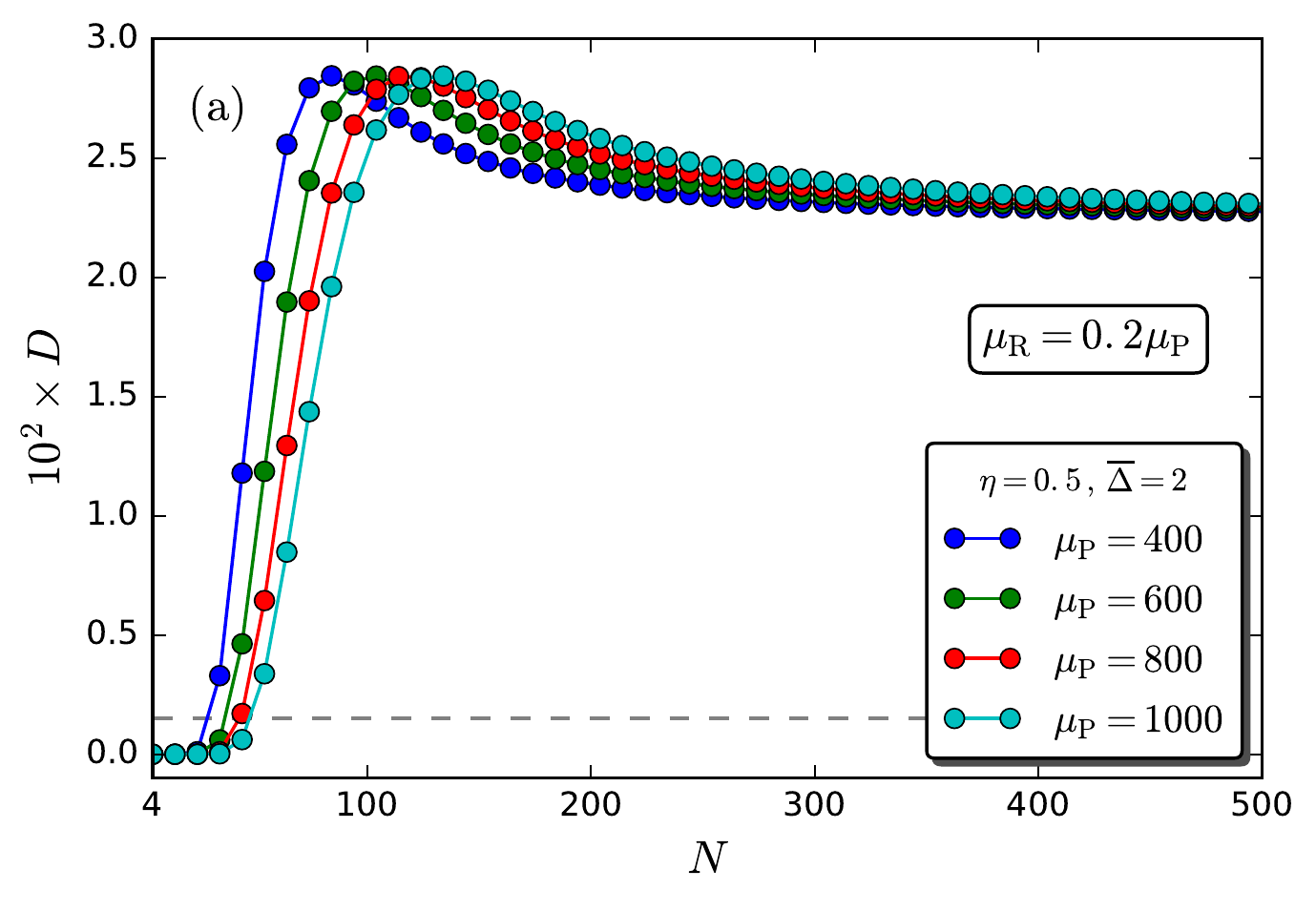}
		\includegraphics[width=0.49\textwidth]{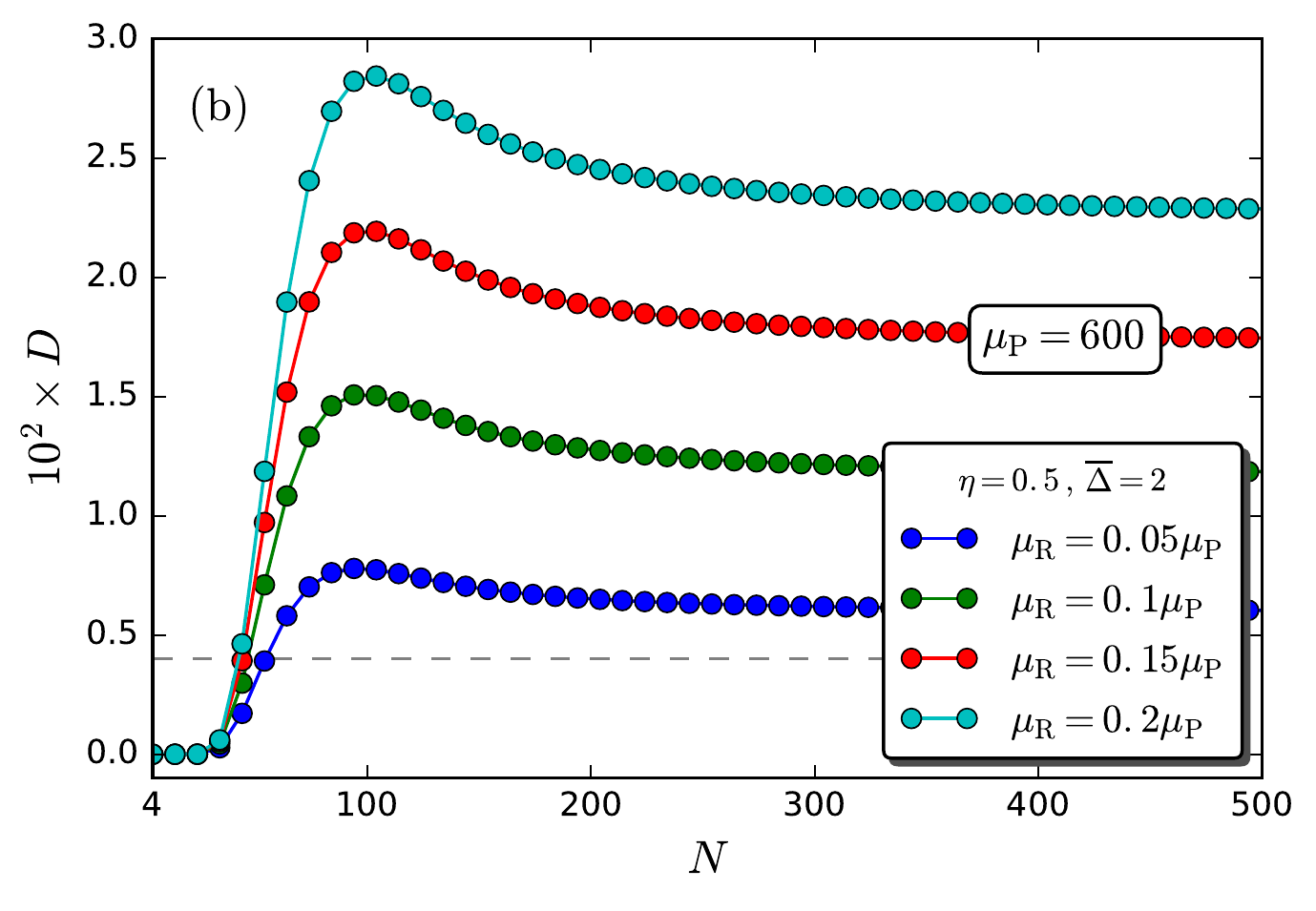}\\
		\includegraphics[width=0.49\textwidth]{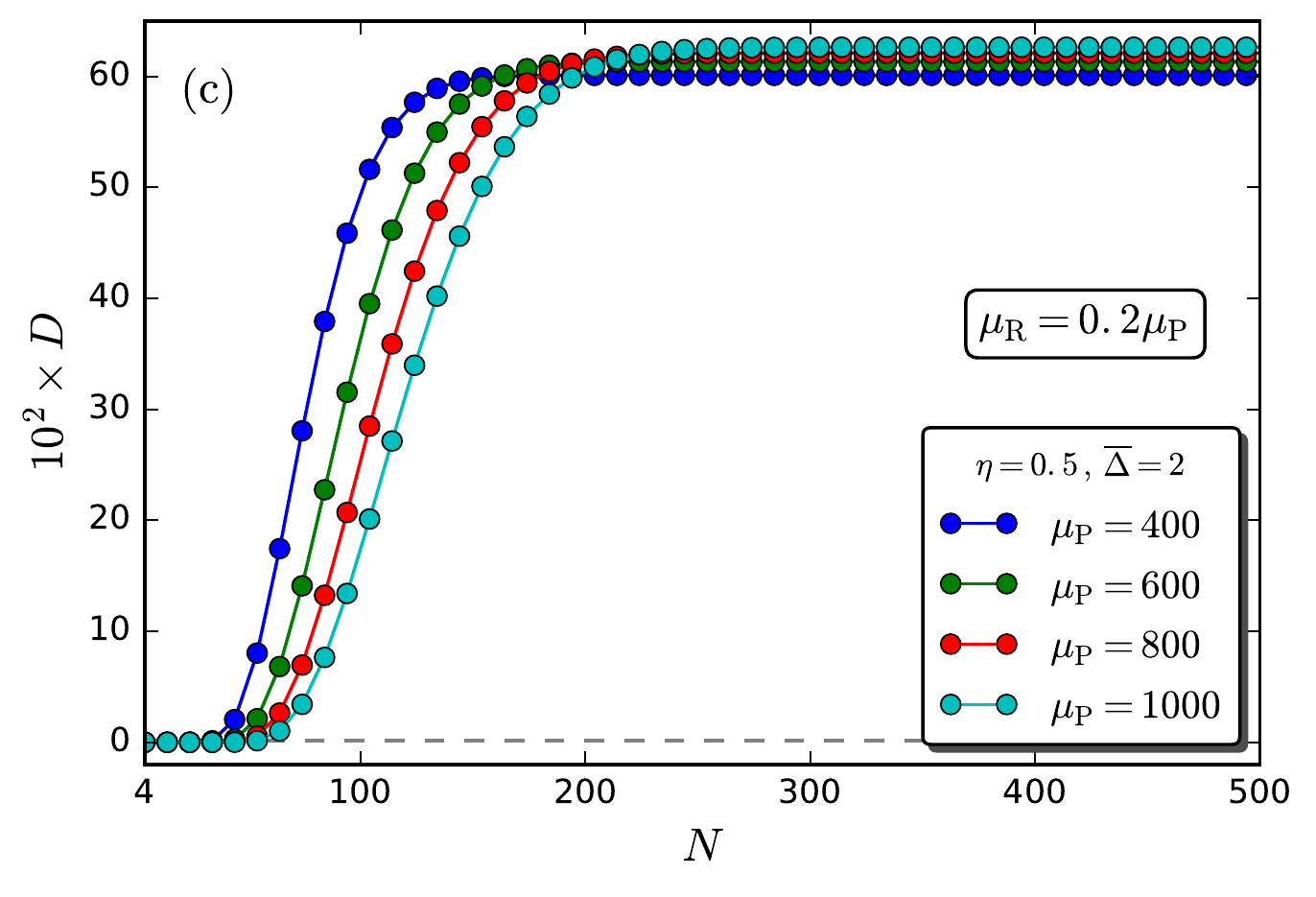}
		\includegraphics[width=0.49\textwidth]{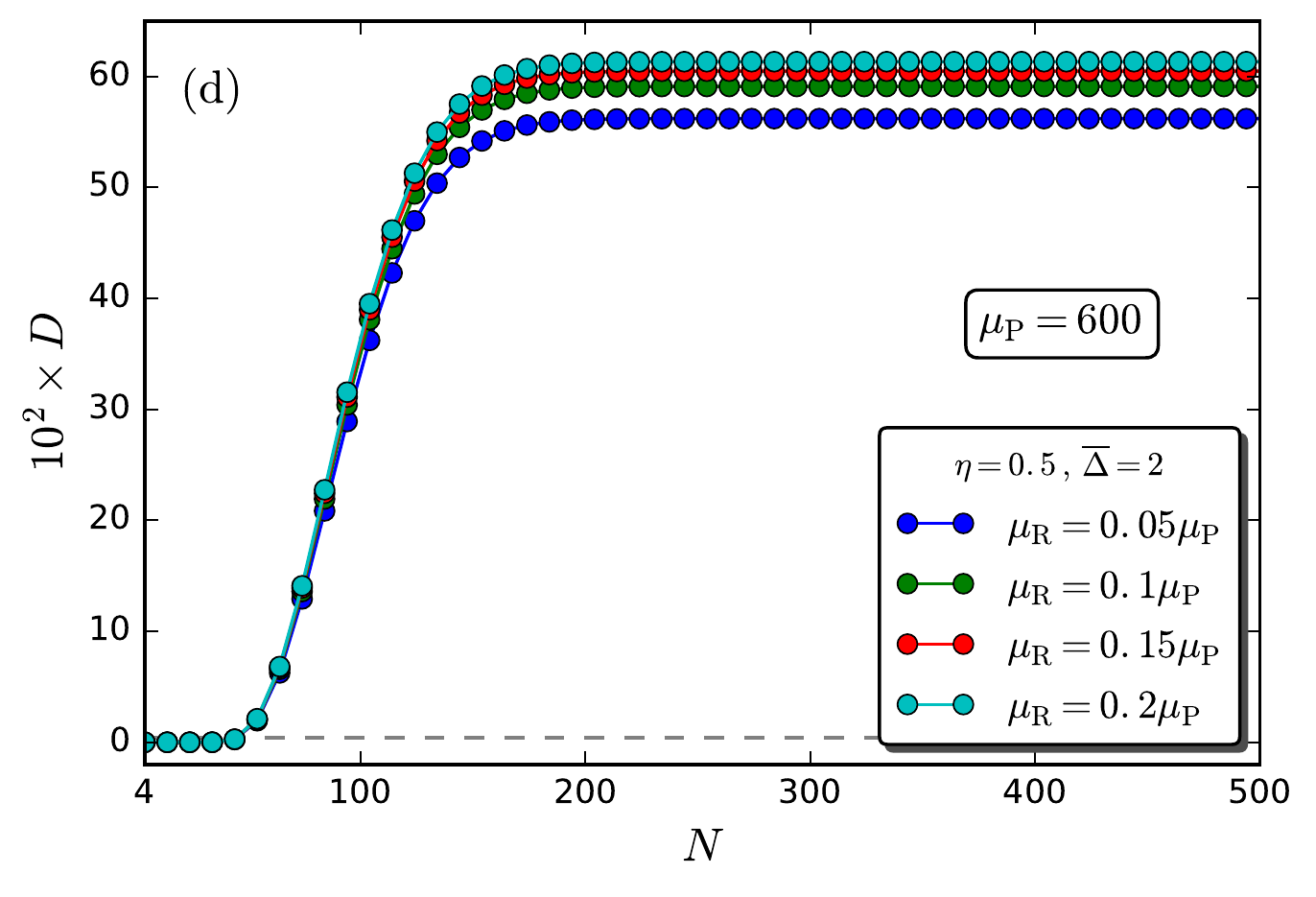}\\
			\includegraphics[width=0.49\textwidth]{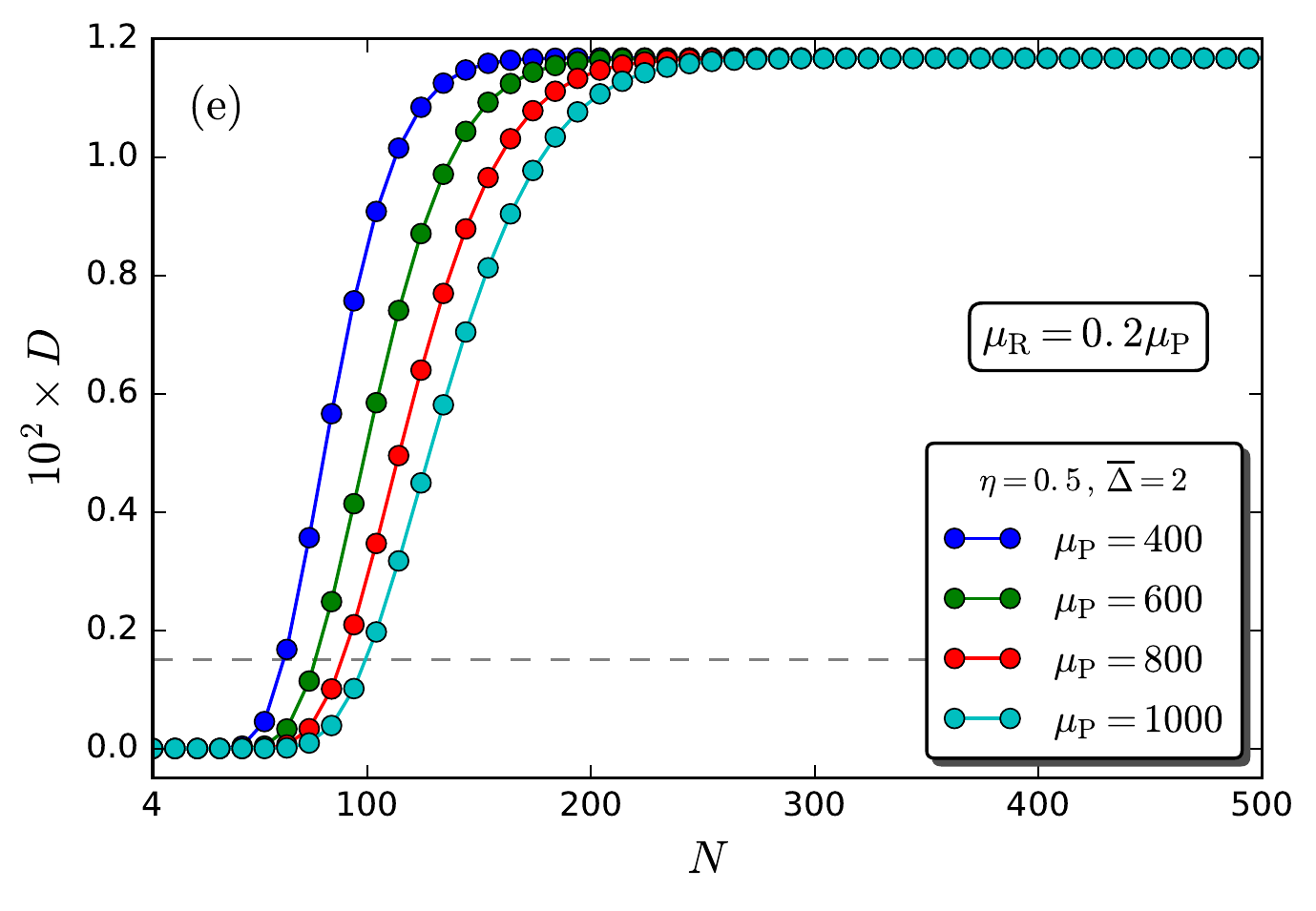}
		\includegraphics[width=0.49\textwidth]{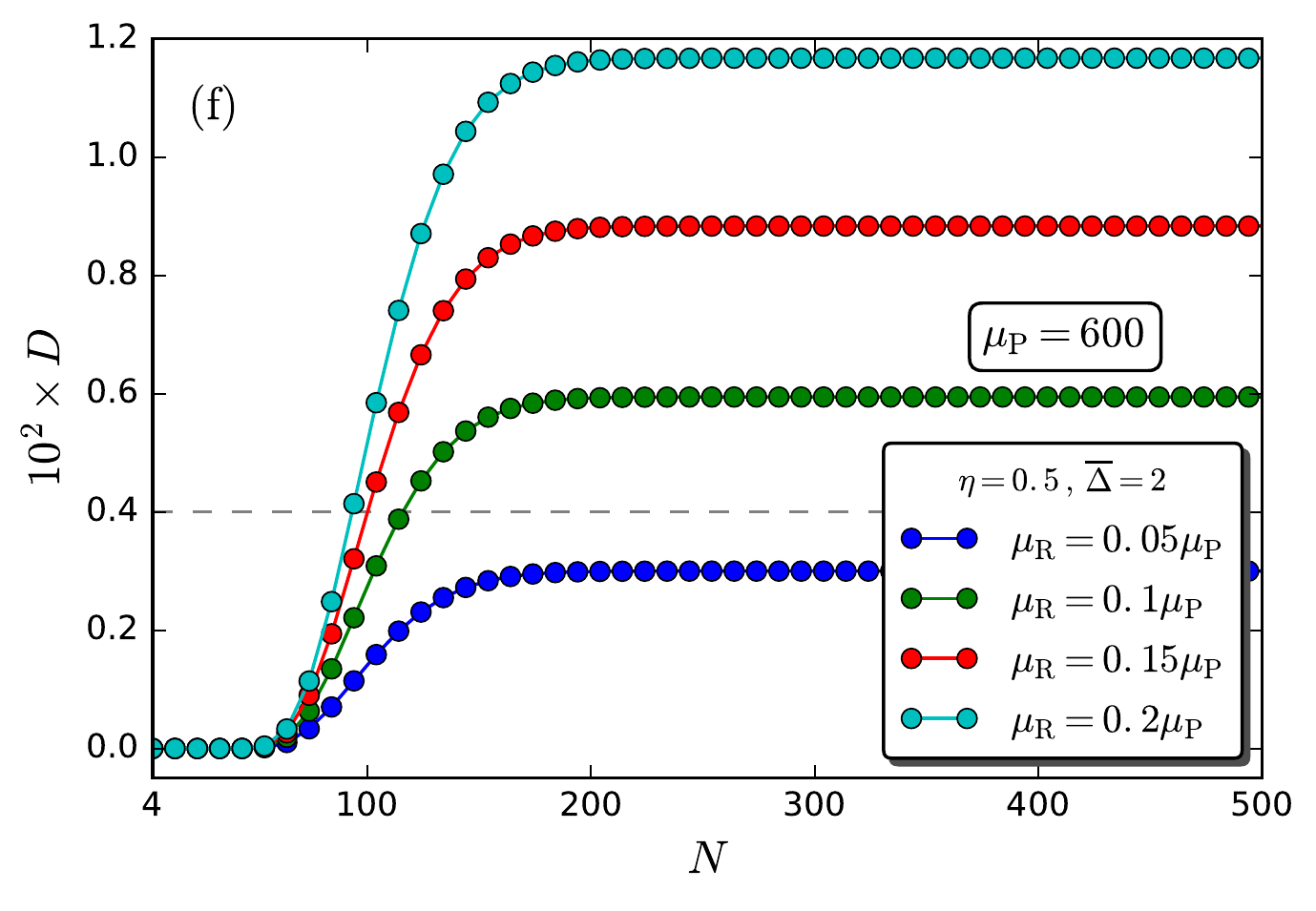}
		\caption{Difference of probabilities 
		$D=P^0_\text{in}-P_\text{in}$ as a function of the total number of different probe states $N$, for the dual-homodyne-detection attack (a,b); the unambiguous-state-discrimination attack (c,d); and the square-root-measurement attack. The horizontal dashed lines mark the security threshold $2\varepsilon = 15\times 10^{-4}$ (l.h.s) and $2\varepsilon = 4\times 10^{-4}$ (r.h.s). The protocol is secure against a specific type of intercept-resend attack when the corresponding value of $D$ exceeds $2\varepsilon$.}
		\label{AllD:fig}
	\end{figure*}

Based on these findings, we expect the lower bound 
 $D_{\rm low}$ to approach zero as we increase $N$ (which is in accordance with Fig. \ref{DLow:fig}), whereas for all of the three attacks,  $D$ is expected to  approach a non-zero value as we increase $N$. 
 Indeed, as shown in Fig. \ref{AllD:fig}, this is confirmed by our simulations.  
As a result, for a given security threshold $\varepsilon$ and fixed losses 
(i.e., fixed ratio $\mu_{\rm R}/\mu_{\rm P}$), there is a very 
broad regime of values for $N$ and $\mu_{\rm P}$ where $D$ exceeds $2\varepsilon$  and the protocol is secure against all of the three attacks.  The secure regime is considerably broader than the one predicted by the lower bound (compare to Fig. \ref{DLow:fig}). The strongest attack seems to be the square-root-measurement attack. Consider, for instance, the case of $2\varepsilon =15\times 10^{-4} $.  We have $D^{\rm (SR)}>2\varepsilon$ for $N> 100$ [see Fig. \ref{AllD:fig}(e)], whereas 
for the other two attacks we find $D^{\rm (DH)}>2\varepsilon$ for $N> 45$ 
[see Fig. \ref{AllD:fig}(a)], $D^{\rm (UD)}>2\varepsilon$ for $N> 60$ [see Fig. \ref{AllD:fig}(c)]. By contrast to the lower bound $D_{\rm low}$ [see Fig. \ref{DLow:fig}(a)], where there is an upper bound on the values of $N$ for which the condition $D_{\rm low}>2\varepsilon$ is satisfied, we do not find any such bound for either of the three attacks we have considered in this work.  Moreover, for all the three attacks we find that the rise of $D$ with $N$ gets  slower as we increase $\mu_{\rm P}$, which is in agreement with the analogous behavior obtained for the error probability.

Keeping the mean number of photons  $\mu_{\rm P}$ in the probe constant, and increasing the losses in the set-up (i.e., decreasing the mean number of photons that reach the HD set-up relative to $\mu_{\rm P}$), we find that lower security thresholds have to be attained by the verifier, so that to ensure the security of the protocol against both of the dual-homodyne-detection and the square-root-measurement attack. For example, we see that the protocol is not secure against the square-root-measurement attack when $2\varepsilon=4\times 10^{-4}$ and $ \mu_{\rm P}=600$, with $\mu_{\rm R} = 0.05\mu_{\rm P}$, but it is secure against the dual-homodyne-detection attack if $N>50$ [see Figs. \ref{AllD:fig}(f) and (b), respectively]. Moreover, for the particular chosen value of $2\varepsilon$, we find that the protocol becomes secure against both of these attacks  if $\mu_{\rm R} = 0.1\mu_{\rm P}$ 
and $N>110$. Security against the two attacks when the losses in the set-up are such that $\mu_{\rm R} = 0.05\mu_{\rm P}$, requires $N\geq 110$ and sufficiently large sample size $M$ so that $2\varepsilon\lesssim 3\times 10^{-4}$.  By contrast, we find small changes in the variation of 
$D^{\rm (UD)}$ with $N$, which suggests that the unambiguous-state-discrimination attack  is weaker than the other two.

\section{Discussion}
\label{sec5}

	In this work we have analysed the seurity of the verification stage in the EAP of Ref. \cite{nikolopoulos2017continuous} against three specific  intercept-resend emulation attacks.  Our results suggest that the protocol is secure against all of the attacks simultaneosuly for a broad range of values for the relevant parameters of the protocol. Moreover, the performance of all of these attacks has been compared to the lower security bound obtained in Ref. \cite{nikolopoulos2018continuous}, by means of the Fano's inequality and the Holevo bound.  
None of the attacks considered here saturates the expected lower bound. Hence, the existence as well as the details of the attack that will saturate the lower security bound, remain a subject of future work. 
One may wonder, whether the displaced-photon counting \cite{dpc2016} can achieve this goal. We believe that this is not the case for a number of reasons. Firstly, the probability for the adversary to deduce the wrong state is expected to be higher for  the displaced-photon counting, than for the square-root measurement discussed here. 
This is because our protocol relies on a set of symmetric states, and it is well-known that the square-root measurement minimizes the probability of error in this case \cite{barnett2009quantum,weir2018optimal}. Higher error probability means that it is easier for the verifier to detect the adversary's intervention. Secondly, the  displaced-photon counting has been shown to outperform the dual-homodyne measurement in phase estimation, only for  phases in a narrow region around $\varphi=0$ \cite{dpc2016}. Outside this region, the dual-homodyne measurement outperforms by the same amount the displaced-photon counting. In our protocol, the random phase of a probe state lies 
in the interval $[0,2\pi)$, and all of the possible values are equally probable. As a result, on the average, the performance of the displaced-photon counting is expected to be very close to the performance of the dual-homodyne detection discussed here.

As far as experimental realisations of the present attacks are concerned,  the dual-homodyne-detection attack can be implemented with current technology \cite{DHDexp}. On the contrary, to the best of our knowledge, the experimental realization of the unambiguous-state-discrimination attack and of the square-root measurement attack for arbitrary number of symmetric coherent states is not known. A rather simple linear-optics implementation of unambiguous state-discrimination has been proposed in Ref. \cite{van2002unambiguous}, which is far from optimal for large values of $N$.  

The present work focuses on the verification of a PUK, assuming that the user has already been verified successfully to the PUK (see Sec. \ref{sec1}). In analogy to conventional smart cards, the verification of the user to the PUK can be achieved through a PIN which is known  to the legitimate owner of the PUK. This is necessary for any EAP, so that to prevent impersonation in  case the PUK is stolen. When the PUK is given to a legitimate user  it is accompanied by the PIN, which has been generated during the enrollment stage, and it is not shown in Fig. \ref{qpuf_scheme:fig}. One way to combine an optical PUK with a PIN is to exploit the techniques of Refs. \cite{Pappu02,Horstmayer13}. For example,  one can assume that the PUK is illuminated by 
classical light, whose wavefront has been modified by a random SLM phase mask ${\bm \Psi}$ [The pattern ${\bm \Psi}$ should not be confused with the optimal phase masks ${\bm \Phi}$ (and ${\bm \Phi}_k$), which are used for the verification of the PUK as discussed in the previous sections of the present work. The former is a totally random phase mask, whereas the latter are optimized with respect to a particular target mode at the output.].  
The response of the PUK to classical light with random wavefront ${\bm \Psi}$ is a random speckle, which can be processed with standard algorithms to result in a numerical  random binary string, say ${\bm w}= {\bm w}_{\rm id}||{\bm w}_{\rm pin}$ \cite{Horstmayer13}. The last $n$ bits of ${\bm w}$,  denoted by ${\bm w}_{\rm pin}$, define the $n-$bit PIN, whereas the first part ${\bm w}_{\rm id}$ serves as an identification number for the list of CRPs to be used for the authentication of the particular PUK (see table \ref{tab1}).  In this way, the list of CRPs, the PUK and the PIN are linked through a random phase mask, which is known only to the verifier, as well as to the authority that issues and distributes the PUKs. 

In analogy to conventional smart cards, at the beginning of the verification stage the user types in his PIN, and  inserts his PUK to the verifier's set-up. The verifier imprints  the pattern  ${\bm \Psi}$ on the wavefront of the incoming light by means of his SLM, and the speckle of the scattered light is processed to yield the classical string ${\bm w}$. If the last $n$ bits of ${\bm w}$ agree with the PIN of the user, the verifier proceeds to the verification of the PUK, otherwise the authentication is aborted immediately. For the verification stage, the verifier sends to the server the key 
$ {\bm w}_{\rm id}$ over a secure and authenticated classical  channel, and the server returns a sequence of $M$ challenges chosen at random from the list of CRPs with identification number equal to $ {\bm w}_{\rm id}$. The verification stage proceeds as discussed in Sec. \ref{sec2}, as well as in Refs. \cite{nikolopoulos2017continuous,nikolopoulos2018continuous}.  
If an adversary has stolen the PUK, he cannot impersonate the legitimate owner without knowledge of the PIN, while the probability for an adversary to guess correctly the $n-$bit PIN is negligible. Moreover, the random string ${\bm w}$ 
is essentially the output of a physical one-way function with input 
${\bm \Psi}$, while the probability for the $j$th bit of ${\bm w}$ to be 0 or 1, is expected to be approximately equal to 0.5 \cite{Pappu02,Horstmayer13}. As a result,  it is hard for an adversary to infer the  random pattern ${\bm \Psi}$, or the PIN from ${\bm w}_{\rm id}$. 
It is worth emphasizing that one can incorporate both of the validation of the PIN and the verification of the PUK in the set-up of Fig. \ref{setup:fig}, after a few amendments. 
More precisely, by adding a half-wave plate and an additional PBS at the output, one can select to image  the speckle onto a camera (for the PIN validation), or onto the optical plane where  SMF B is positioned (for the PUK verification)   \cite{huisman2014controlling,defienne2014nonclassical}. 
Hence, the set-up can operate in two different modes. Note also that in the "PIN-validation"  mode, intense probe has to be used so that to obtain a clear speckle, whereas in the "PUK-verification" mode the set-up has to operate with parameters that 
are dictated by the present as well as previous security analyses. 
Finally, we cannot exclude alternative ways to link an optical PUK to a PIN e.g., through the control of laser parameters (such as the angle of incidence and the wavelength) \cite{Pappu02}, and/or the control of the target mode at the output \cite{Nikolopoulos:19}.

\section {Acknowledgements} 
This work has been funded by the Deutsche Forschungsgemeinschaft as part of the 
project S4 within CRC 1119 CROSSING. 
L.F. and G.M.N. are grateful to Prof. W. H. Pinkse for his hospitality and enlightening discussions, as well as to  Prof. S. M. Barnett for useful suggestions 
related to  state discrimination.


\end{document}